
\magnification 1200
\baselineskip=20pt
\centerline{{\bf Unified QCD evolution equations and the dominant behaviour}}

\centerline{{\bf  of structure functions at low $
 x $}}

\centerline{by}

\centerline{R. Peschanski and S. Wallon}

\centerline{Service de Physique Th\'eorique, CEA-Saclay}
\centerline{F-91191 Gif-sur-Yvette Cedex, FRANCE}
\vglue 2truecm
\centerline{{\bf ABSTRACT}}

We consider a system of evolution equations for quark and gluon
structure functions satisfying the leading-logarithmic
behaviour due to both QCD collinear $ \left(LLQ^2 \right) $ and infrared $
(LL1/x) $
singularities.
We show that these equations leave undetermined an arbitrary regular function
of $j$ in the Mellin-transformed weights.
We consider the constraints resulting from
energy-momentum
conservation and from the decoupling of quark loops in the leading $ j $-plane
singularity. These constraints can be fulfilled without influencing the
leading-log terms. As a
particular consequence of the second constraint, the location of the leading
singularity is determined in terms of
the $ (LL1/x) $ and $ \left(LLQ^2 \right) $ kernels.
It leads to a value
significantly lower than the $ LL1/x $ evaluation, while remaining
at $j > 1,$ and compatible with the behaviour of structure
functions observed at HERA.
\vfill\eject

The recent results on quark and gluon structure functions at Hera have
paved the way for a reconsideration of QCD
predictions.
 The observed behaviour of
the
proton structure function $ F_2 \left(x,Q^2 \right) $ and some indications on
the gluon
structure function $ F_G \left(x,Q^2 \right) $ in the range $ \left\{
10^{-4}\leq x\leq 10^{-2}, 8\leq Q^2\leq 60\ \rm {GeV}^2 \right\} $ are
characterized$ ^{[1]} $ by a rapid rise at small $ x $ which is
qualitatively compatible
 with the predictions of the resummation of leading $\log (1/x)$
 contributions (hereafter denoted $ LL1/x $) of the
perturbative expansion, i.e. with the Lipatov singularity$ ^{[2]}$ (BFKL).
There exists quantitative studies of structure functions including the BFKL
singularity$ ^{[3]}.$

 On the
other
hand, new tests of the celebrated
Altarelli-Parisi evolution equations$ ^{[4]} $ (DGLAP) are now possible
in a much larger $ Q^2 $ range
.  These equations
correspond to the resummation of the leading
 log($Q^2$) terms (denoted $LLQ^2$)
of the perturbative QCD expansion.
Thus, HERA represents a unique apparatus for testing the
QCD theoretical tools in kinematical domains where the perturbation expansion
has
to be resummed. As well-known, this is due to the appearence of
collinear and infrared (in the infinite-momentum frame) singularities in the
perturbative theory, implying
large logarithms in the effective coupling constant. Hopefully, a
precise comparison between
experiment and theory will help understanding
the yet unknown non-perturbative regime.
In this context, a unified
description of the $ LL1/x
$
and $ LLQ^2 $ evolution equations for quark and gluon structure functions
is highly desirable.

Two different approaches
have already been proposed, each one with its own advantages
and inconveniences. First, it was remarked$ ^{[5]} $ that Feynman diagrams for
multi-gluon
emission which contribute to $ F_G \left(x,Q^2 \right) $ are
characterized by a common
angular ordering property in the whole $ x $-range, leading after resummation
to the BFKL singularity at small $x$ and to the DGLAP equations elsewhere. This
allows one to write
a unique equation for the whole $x$ range$ ^{[6]}. $
This approach leads to fruitful Monte Carlo simulations$ ^{[7]} $ based
on the multi-gluon diagrams but
an explicit solution of the evolution equation$ ^{[6]}$ itself
has not yet been found.

In a second existing approach, explicit solutions have been proposed in a
different context$ ^{[8]}$.
Using a constructive procedure starting directly from the structure functions,
a system of evolution equations can be written following the scheme:
$$
(LL1/x)+ \left(LLQ^2 \right)-(DLL), \eqno (1)
$$
which indicates that the sum of $ LL1/x $ and $ LLQ^2 $ contributions is
corrected for
double-counting by substraction of the double-leading-logarithmic terms $
(DLL). $
This
led the authors$^{[8]}$  to propose a set of explicit equations for the
quark-gluon system.
However, as we shall see below, we find that the realization of the scheme (1)
in terms of an explicit set of evolution equations leads to a solution which
differs from that proposed in Ref.$ [8].$ On the one hand, the set of equations
obtained there leaves undetermined a regular function related to higher order
contributions. On the other hand, the realization of (1) has to obey specific
constraints which should be satisfied by the resulting equations, namely i)
energy-momentum conservation, and ii) the decoupling of quark loops in the BFKL
kernel. These
constraints are not satisfied by the system of Ref.$ [8].$

The goal of this paper is to provide
an  explicit realization of the formal scheme (1), obeying the
constraints. We thus provide an explicit
solution of the fundamental equation written in Ref.$ [6], $ while extending
it to the quark-gluon system.

The generic set of equations we propose following the scheme (1) reads:

$$
 \eqalignno{{ {\rm d} F_G \left(j,Q^2 \right) \over {\rm d} \ln Q^2}
&\equiv Q^2 f_G
\left(j,Q^2 \right)=Q^2_0 f_G \left(j,Q^2_0 \right)+ \int^{
Q^2}_{Q^2_0}{\alpha_ S \left(Q^{\prime 2} \right) \over 4\pi} {\rm d}
Q^{\prime 2} & \cr
&  \left[\left\{ \left(\nu_ G+ \Psi \right)(j)\ Q^2K
\left(Q^2,Q^{\prime 2} \right)-\Psi (j) \right\} f_G \left(j,Q^{\prime 2}
\right)+\phi^ G_F(j)f_S \left(j,Q^{\prime 2} \right) \right] &
 \cr
{ {\rm
d} F_S \left(j,Q^2 \right) \over {\rm d}\ln Q^2}&\equiv Q^2 f_S \left(j,Q^2
\right)=Q^2_0 f_S \left(j,Q^2_0 \right)+ \int^{ Q^2}_{Q^2_0}{\alpha_ S
\left(Q^{\prime 2} \right) \over 4\pi} {\rm d} Q^{\prime 2}  & \cr
&\left[\nu_ F(j)f_S \left(j,Q^{\prime 2} \right)+2n_ F\phi^ F_G(j)f_G
\left(j,Q^{\prime 2} \right) \right], & (2) \cr}
 $$
where the inhomogeneous terms of the system of equations is implied
by the compatibility with the DGLAP evolution equations near the
threshold $Q^2_0.$ $ \Psi (j) $ is an arbitrary  function
provided it remains  regular at $
j=1,$
and where
$$ \alpha_ S \left(Q^2 \right)= 1 / (b\  {\rm ln}\ Q^2/\Lambda^
2) ;\
\left\{ \nu_ G,\phi^ G_F,\nu_
F,\phi^ F_G \right\}(j),  $$
are respectively  the first-order QCD coupling constant, and
the set of Altarelli-Parisi weights written with the
usual conventions$ ^{[9]}. $ Here $
K \left(Q^2,Q^{\prime 2} \right) $
is the BFKL kernel $ ^{[2]}.$
Note
that the specific equations of Ref.$ [7] $ are recovered with the
choice $ \Psi
= 4N_c /(j-1)-\nu_ G. $
As we shall now demonstrate, this choice is not required by (1)
and moreover it contradicts energy-momentum conservation and the decoupling of
quark loops in the BFKL kernel.

Let
us briefly show how the system (2) verifies both $LL1/x \ {\rm and}\ LLQ^2$
behaviour,
 while leaving undetermined the regular function $ \Psi (j). $
It is well known$ ^{[8]} $ that the kernel $ K \left(Q^2,Q^{\prime 2} \right)
$ reduces simply to $ 1/Q^2 $ when is retained only its contribution to the
$LLQ^2$ terms.
 In that precise case, it is easy to realize by substitution in (2)
that the function $ \Psi (j) $ disappears from the system.
Integrating by part Eqns.(2),
one
recovers the ordinary DGLAP evolution equations
up to negligible terms.

The $ LL1/x $ reduction of the system (2) is recovered by selecting the
most singular contributions in the complex $j$-plane. This is obtained first
by considering the equation with only the BFKL kernel $ K \left(Q^2,Q^{\prime
2} \right).
$
Then one has also to select the most singular contribution from the prefactor
$ \nu_ G+\Psi \approx 4N_c /
(j-1).$ By this way, one finds back the usual BFKL kernel.
Again, the function $ \Psi (j) $ remains undetermined, provided that it is
chosen regular at $ j=1. $
By this discussion, we have proved that the requirement of
equation (1) has a large class of solutions
parametrized by  the regular but arbitrary function $ \Psi (j). $

Now, let us introduce the constraints on the system of equations (2) which
cannot be satisfied by the leading-log terms only . The
first one is due to energy-momentum conservation  which reads:
$$ \left.\forall Q^2, \left(F_G \left(j,Q^2 \right)+F_S \left(j,Q^2 \right)
\right) \right\vert_{ j=2}=1 \Longrightarrow f_G \left(2,Q^2 \right)+f_S
\left(2,Q^2 \right)=0.\eqno (3) $$
Inserting the constraint (3) (for both $ Q^2 $ and $ Q^2_0) $ in the system (2)
taken at $ j=2, $
one gets:
$$ 0\equiv \int^{ Q^2}_{Q^2_0}{\alpha_ S \left(Q^{\prime 2} \right) {\rm d}
Q^{\prime 2} \over 4\pi} \left(\Psi (2)+\nu_ G(2) \right) \left( K
\left(Q^2,Q^{\prime 2} \right)-1/Q^2 \right),\eqno (4) $$
where one uses the values of the Altarelli-Parisi kernels at $ j=2, $ namely$$
\left. \left\{ \nu_ G,\phi^ G_F,\nu_ F,\phi^ F_G \right\} \right\vert_{
j=2}\equiv \left\{ -{2 \over 3}n_F,{8 \over 3}C_2,-{8 \over 3}C_2,{1 \over 3}
\right\} . $$
Then, if the kernel $ K \left(Q^2,Q^{\prime 2} \right) $ reduces to $1/Q^2$
 the identity (4)
is trivially fulfilled. Since it is not the case beyond the double-leading-log
approximation, energy-momentum conservation requires the relation $
\Psi (2)={2 \over 3}n_F. $
This is the first condition on the
previously undetermined function $ \Psi (j). $

Let us now consider solutions of the system (2). In the present paper, we shall
restrict
ourselves to the conventional case of a fixed coupling constant $ \alpha_
S\equiv\bar
\alpha . $
Indeed, this is the only case for which
the BFKL singularity has been derived from the diagrammatic expansion$ ^{[2]}.
$
Following a  common procedure, usual in the $LL1/x$ case$ ^{[2,8,10]}, $ we
 write the equations for $ f_{G,S}$ in terms of their
inverse
Mellin transforms, namely:
$$ f_{G,S} \left(j,Q^2 \right)\equiv \int^{ }_ C{ {\rm d} \gamma \over 2i\pi}
\ \varphi_{ G,S}(j,\gamma ) \left(Q^2 \right)^{\gamma -1}, \eqno (5) $$
where $ C $ is a contour in the
complex-$ \gamma $
plane leaving all singularities of the integrand at its left. Inserting the
Mellin transform
(5) in the system (2), and after some manipulations, one finds
the following matrix equation:

$$ \left( \matrix{ 1-{\bar \alpha \over 4\pi} \left\{ \left(\nu_ G+\Psi
\right)\omega (\gamma )-{\Psi  \over \gamma} \right\} \hfill&
&
 -{\bar
\alpha \over 4\pi}
{ \phi^ G_F \over \gamma} \cr  \hfill&    &   \cr
 -{\bar
\alpha \over 4\pi}
{2n_F\phi^ F_G \over  \gamma} \hfill&    &  1-{\bar \alpha
\over 4\pi}{ \nu_ F \over \gamma} \cr} \right) \left( \matrix{ \varphi_
G \cr  \cr \varphi_ S \cr} \right)= \left( \matrix{
\varphi_{ 0G} \cr  \cr \varphi_{ 0S} \cr} \right) \eqno
(6)
$$

\noindent where $ \omega (\gamma )= 2\psi(1) - \psi(\gamma) - \psi(1-\gamma),$
with $\psi(\gamma) = {{\rm d} \ln \Gamma \over {\rm d} \gamma},$ is the
well-known eigenvalue of
the BFKL kernel\footnote{$ ^1 $}{The scale dependence of the system
has been merged into the right-hand side of eq.(6),
since it is not
relevant for the computation of the BFKL singularity at fixed
coupling. However a treatment
with a running coupling constant would imply to take into account the
threshold $ Q^2_0, $ see for instance Ref.$ [10]. $}. Note that the vector of
the right-hand side (6)
includes the boundary-value
conditions on the
system (2). It is  usually assumed not to contain the dominant $ j $-plane
singularity.

By matrix inversion one gets:

$$
\left( \matrix{ \varphi_ G \cr  \cr \varphi_ S \cr}
\right)={1 \over {\cal D}(j,\gamma)}
\cdot \left( \matrix{ 1-{\bar \alpha \over 4\pi \gamma} \nu_ F  &    & {\bar
\alpha \over 4\pi} \phi^ G_F \cr   &    &   \cr{\bar \alpha \over 4\pi}
2n_F\phi^ F_G  &    & 1-{\bar \alpha \over 4\pi} \left\{ \omega (\gamma )
\left(\nu_ G+\Psi \right)-{\Psi \over \gamma} \right\} \cr} \right) \left(
\matrix{ \varphi_{ 0G} \cr  \cr \varphi_{ 0F} \cr} \right), \eqno (7) $$
where
$$
{\cal D} (j,\gamma)=
1-{\bar \alpha \over 4\pi} \left\{ \omega (\gamma )
\left(\nu_ G+\Psi \right)+{1 \over \gamma} \left(\nu_ F-\Psi \right) \right\},
$$
neglecting contributions of order $ \bar \alpha^ 2. $

In order to obtain the leading behaviour
of structure functions at low $x,$ we  perform a double
inverse-Mellin
transform on the system (7), first in the variable $ j, $ then in the variable
$ \gamma:$
$$ f_{ G,S} \left(x,Q^2 \right)\equiv \int^{ }_{ }{ {\rm d} \gamma \over
2i\pi} {\rm e}^{(\gamma -1) {\rm ln}  Q^2} \int^{ }_{ }{ {\rm d} j \over
2i\pi} {\rm e}^{(j-1) {\rm ln}  1/x}\ \varphi_{ G,S}(j,\gamma ). \eqno (8) $$
 The first integral of Eq. (8) (in the complex j-plane) is given by the residue
of the pole at  $ j = j_p(\gamma ) $ where ${\cal D} (j_p,\gamma) = 0.$ Then
the second Mellin transform (in the complex $\gamma$-plane) is performed by a
saddle-point method. In the domain of validity of the BFKL singularity, one
assumes ${{\rm ln} Q^2 \over{\rm ln} 1/x} \ll 1,$ and the saddle-point is at  $
\gamma =\gamma_ c $ corresponding to ${{\rm d} j_p (\gamma_c)  \over {\rm d
}\gamma} = 0.$
One finds:
$$ (pole) \ \ \ {4\pi \over\bar \alpha} = \left(\nu_ G \left(j_p
\right)+\Psi \left(j_p \right) \right)\omega (\gamma )+ \left(\nu_ F \left(j_p
\right)-\Psi \left(j_p \right) \right){1 \over \gamma} \eqno  $$
$$ (saddle\ point) \ \ \ 0= \left(\nu_ G \left(\bar j
 \right)+\Psi \left(\bar j \right) \right)\omega^{
\prime} \left(\gamma_ c \right)-{1 \over \gamma^ 2_c} \left(\nu_ F \left(\bar j
 \right)-\Psi \left(\bar j \right)
\right), \eqno (9) $$
where $ \bar j = j_p(\gamma_c).$
The resulting structure functions behave like:
$$ f_{ G,S} \left(x,Q^2 \right)\approx \left[Q^2 \right]^{\gamma _c-1}x^{-
\left(\bar j-1 \right)}, \eqno (10) $$

Now, we are able to express the second constraint on the system of equations
(2), namely the decoupling of quark loops from the BFKL singularity kernel. Let
us first require that the dominant behaviour (10) agrees with the theoretical
determination of the BFKL
singularity$^{[2,8]}$. It corresponds to the well-known \lq\lq critical\rq\rq\
value of $
\gamma, $ namely $ \gamma_ c=1/2, $ with $ \omega^{ \prime} \left(\gamma_ c
\right)=0 $ and $\omega(\gamma_ c) = 4 \ln 2.$
{}From inspection of the system (9), this implies the following relations:
$$ \eqalignno{ \Psi \left(\bar j \right) & =\nu_ F \left(\bar j \right) &  \cr
\nu_ G \left(\bar j \right)+\nu_ F \left(\bar j \right) & = \left[{4\bar
\alpha \ {\rm log} \ 2 \over 4\pi} \right]^{-1}. & (11) \cr} $$

We note that the first of relations (11) implies the cancellation of the quark
contribution $\nu_F - \Psi$ in the pole location, see (7). Indeed, it is
known$^{[2]}$ that a loop contribution for a particule of spin $s$ is of order
$x^{-(2 s - 1)}$ at small $x.$ Thus, quark loops are negligeable in the BFKL
kernel.

The results of Eqs.(11) have some interesting consequences on the location of
the BFKL singularity in the $j$-plane.

 If we retain only the dominant behaviour $ \nu_ G  \approx  { 4N
\over j-1}, $ when $j\mapsto 1 ,$ we recover the usual determination of the
BFKL singularity, in the absence of $LLQ^2$ contributions.
The coupling to the $ LLQ^2 $ terms, which is allowed by our unified
description, is responsible for the
modified
formula (11).
As an illustration, let us for instance fix the value of $ \bar
\alpha $ such that $ {N_c \over \pi}\bar \alpha \ 4\ {\rm ln} \ 2\simeq .5. $
This value is often considered as fixing the \lq\lq bare\rq\rq\ BFKL
singularity$^{[3]}$. Using this
value and fixing $ N_c=n_F=3 $, one gets from the second
equation (11): $ \bar j-1=0,313. $
Quite interestingly,
this
value is in better agreement with those phenomenologically
obtained from Hera experimental results, if analyzed directly in terms of a
BFKL singularity$^{[1]}.$ Note that, with these input
numbers,
one gets for the function $ \Psi $ the two constraints $ \Psi (\bar j)\equiv
\nu_ F(j=1.313)\simeq -1.44; $
$ \Psi (2)\equiv \nu_ G(2)=2. $
The kinematical constraint of energy-momentum conservation decreases the
intercept of the BFKL
singularity$^{[11]},$ through non-leading terms in the $LL1/x$ expansion.
However, this constraint at $j = 2$ does not allow a determination of the
modified location of the singularity.
 In our derivation, see equation (11), we obtain a precise theoretical
prediction for this shift, based on a different constraint, e.g. the decoupling
of the quark loops from the BFKL kernel.

Note that in all realistic cases (values of $n_F$, flavour thresholds, etc...),
$  \nu_ G \left(j \right)+\nu_ F \left(j \right)  < 4 N /(j-1) ,$ for $j>1.$
Thus the decoupling constraint always decreases
the BFKL intercept $\bar j - 1.$

In conclusion, we have proposed QCD evolution equations combining
the
leading-logarithmic contributions in both the $ x $ and $ Q^2 $ variables and
satisfying constraints, namely
energy-momentum conservation and quark decoupling from
the BFKL
singularity. Our results extend the QCD
predictions
beyond the leading-logarithm contributions. In particular, we find the
location of the dominant singularity of structure functions at small-$ x $ and
fixed
coupling constant, which is shifted down with respect to the $
LL1/x $
prediction. Other consequences of this improved leading-logarithm scheme should
be explored. In particular, it would be interesting to see in which way the
system
(2), with an adequate function $ \Psi (j) $, gives an explicit realization of
the
gluon resummation of Ref.$ [5] .$
\vskip 20mm
\noindent {\bf ACKNOWLEDGMENTS}

We would like to acknowledge stimulating discussions with S. Catani, J.
Kwiecinski, E. Levin, G. Marchesini, and
H. Navelet and A. Morel for fruitful remarks.
\vfill\eject
\centerline{{\bf REFERENCES}}
\vskip 20mm
\item{$\lbrack$1$\rbrack$} H$_1$ Coll., {\sl Nucl. Phys.\/} {\bf B407} (1993)
515;
\item{\nobreak\ \nobreak\ \nobreak\ } Zeus Coll., {\sl Phys. Lett.\/} {\bf
B316} (1993) 412,
 and more recent results to
appear in the proceedings of the
Glasgow Conference, July 1994.

\item{$\lbrack$2$\rbrack$} E.A. Kuraev, L.N. Lipatov, V.S. Fadin, {\sl Sov.
Phys. JETP\/} {\bf 45} (1977) 199;
\item{\nobreak\ \nobreak\ \nobreak\ } Ya.Ya. Balitsky and L.N. Lipatov, {\sl
Sov. Nucl. Phys.\/} {\bf 28} (1978) 822.

\item{$\lbrack$3$\rbrack$} See, for instance, A.J. Askew, J. Kwiecinski,
 A.D. Martin, P.J. Sutton, {\sl
Phys. Rev. } {\bf D49} (1994) 4402;

\item{$\lbrack$4$\rbrack$} C. Altarelli and G. Parisi, {\sl Nucl. Phys.\/}
{\bf B126} (1977) 298;
\item{\nobreak\ \nobreak\ \nobreak\ } V.N. Gribov and L.N. Lipatov, {\sl Sov.
Journ. Nucl. Phys.\/} (1972) 438 and
675;
\item{\nobreak\ \nobreak\ \nobreak\ } Yu.L. Dokshitzer, {\sl Sov. Phys.
JETP\/} {\bf 46} (1977) 641.

\item{$\lbrack$5$\rbrack$} M. Ciafaloni, {\sl Nucl. Phys.\/} {\bf B296} (1987)
249;
\item{\nobreak\ \nobreak\ \nobreak\ } S. Catani, F. Fiorani and G. Marchesini,
{\sl Phys. Lett.\/} {\bf B234} (1990) 389 and
{\sl Nucl. Phys.\/} {\bf B336} (1990) 18;
\item{\nobreak\ \nobreak\ \nobreak\ } S. Catani, F. Fiorani, G. Marchesini and
G. Oriani, {\sl Nucl. Phys.\/} {\bf B361}
(1991) 645.

\item{$\lbrack$6$\rbrack$} G. Marchesini, Proc. of Workshop \lq\lq {\it QCD at
200\nobreak\ TeV}\rq\rq , Erice, Italy, June
1990, eds. L. Cifarelli and Yu.L. Dokshitzer, Plenum Press, 1992, p.193.

\item{$\lbrack$7$\rbrack$} G. Marchesini and B.R. Webber, {\sl Nucl. Phys.
\/}{\bf B349} (1991) 617;
\item{\nobreak\ \nobreak\ \nobreak\ } E.M. Levin, G. Marchesini, M.G. Ryskin,
B.R. Webber, {\sl Nucl. Phys.\/} {\bf B357}
(1991) 167.

\item{$\lbrack$8$\rbrack$} L.V. Gribov, E.M. Levin and M.G. Ryskin, {\sl Zh.
Eksp. Teor. Fiz.\/} {\bf 80} (1981)
2132; {\sl Phys. Rep. \/}{\bf 100} (1983) 1.

\item{$\lbrack$9$\rbrack$} See for definitions and notations:
\item{\nobreak\ \nobreak\ \nobreak\ } {\sl Basics of Perturbative QCD\/},
Yu.L. Dokshitzer, V.A. Khoze, A.H. Mueller
and S.I. Troyan (J. Tran Than Van ed., Editions Fronti\`eres) 1991.

\item{$\lbrack$10$\rbrack$} J.C. Collins, J. Kwiecinski, {\sl Nucl. Phys.\/}
{\bf B316} (1989) 307.

\item{$\lbrack$11$\rbrack$} R.K. Ellis, Z. Kunszt, E.M. Levin, {\sl Nucl.
Phys.\/}{\bf\ B420} (1994) 517.

\end